\documentclass{article}

\usepackage{PRIMEarxiv}

\usepackage[utf8]{inputenc} % allow utf-8 input
\usepackage[T1]{fontenc}    % use 8-bit T1 fonts
\usepackage{hyperref}       % hyperlinks
\usepackage{url}            % simple URL typesetting
\usepackage{booktabs}       % professional-quality tables
\usepackage{amsfonts}       % blackboard math symbols
\usepackage{nicefrac}       % compact symbols for 1/2, etc.
\usepackage{microtype}      % microtypography
\usepackage{lipsum}
\usepackage{fancyhdr}       % header
\usepackage{graphicx}       % graphics
\usepackage{bm}
\usepackage{color}
\usepackage{comment}
\usepackage{amsmath}
\usepackage{algorithmic}
\usepackage{algorithm}
\graphicspath{{media/}}     % organize your images and other figures under media/ folder
\usepackage{multirow}
\usepackage{comment}
\usepackage{threeparttable}

\DeclareMathOperator*{\argmin}{arg\,min}

\title{Survival causal rule ensemble method considering the main effect for estimating heterogeneous treatment effects}

\author{
  Ke Wan$^\ast$ \\
  Department of Medicine \\
  Wakayama Medical University \\
  Japan\\
  \texttt{wane19911017@gmail.com} \\
     \And
  Kensuke Tanioka \\
  Department of Biomedical Sciences and Informatics \\
  Doshisha University \\
  Japan\\
  \texttt{ktanioka@mail.doshisha.ac.jp} \\
     \And
  Toshio Shimokawa \\
  Department of Medicine \\
  Wakayama Medical University \\
  Japan\\
  \texttt{toshibow2000@gmail.com} \\
}

\begin{document}
\maketitle

\begin{abstract}
With an increasing focus on precision medicine in medical research, numerous studies have been conducted in recent years to clarify the relationship between treatment effects and patient characteristics. The treatment effects for patients with different characteristics are always heterogeneous, and various heterogeneous treatment effect machine learning estimation methods have been proposed owing to their flexibility and high prediction accuracy. However, most machine learning methods rely on black-box models, preventing direct interpretation of the relationship between patient characteristics and treatment effects. Moreover, most of these studies have focused on continuous or binary outcomes, although survival outcomes are also important in medical research. To address these challenges, we propose a heterogeneous treatment effect estimation method for survival data based on RuleFit, an interpretable machine learning method. Numerical simulation results confirmed that the prediction performance of the proposed method was comparable to that of existing methods. We also applied a dataset from an HIV study, the AIDS Clinical Trials Group Protocol 175 dataset, to illustrate the interpretability of the proposed method using real data. Consequently, the proposed method established an interpretable model with sufficient prediction accuracy.
\end{abstract}

% keywords can be removed

\section{Introduction}
Randomized controlled trials (RCTs) are widely regarded as the gold standard for evaluating treatment effectiveness in evidence-based medicine. In RCTs, the average treatment effect (ATE) can be easily estimated to provide valuable evidence for treatment effectiveness. Although the ATE from RCTs always represents the average effect for specific subject groups, it ignores the heterogeneity of the treatment effect for subjects with various characteristics. Thus, ATEs are unable to provide sufficient information to make optimal treatment decisions for each subject. Therefore, in recent years, estimation of treatment effects has become extremely important across patients with various characteristics \cite{Hill2011,Foster2011,Tian2014,Henderson2016,Athey2016,Wager2018,Sugasawa2019,Cui2023}. However, most of these studies have primarily focused on continuous or binary outcomes; nonetheless, survival outcomes are also crucial in medical research. Therefore, in this study, we focused on estimating heterogeneous treatment effects (HTEs) on survival outcomes. 

Machine learning methods are usually used for existing survival HTE estimation because of their high prediction accuracy \cite{Henderson2016,Sugasawa2019,Cui2023}. However, their reliance on black-box models poses a limitation because they lack explicit insight into the relationships between HTEs and covariates. This lack of interpretability hampers the understanding and trustworthiness of the results, particularly for medical researchers and professionals \cite{Price2018, Petch2022}. To overcome this limitation, we focused on RuleFit, a method known for its interpretability and prediction accuracy, which are similar to those of tree ensemble methods\cite{Friedman2008}. Although RuleFit was initially developed for continuous outcomes, it has been extended to handle survival outcomes based on Cox proportional hazards models \cite{Fokkema2020}. To distinguish it from the original RuleFit, we refer to this approach as the ``survival RuleFit.’’ However, the survival RuleFit can only construct models to interpret the relationships between outcomes and covariates. Therefore, in this study, we propose a survival HTE estimation method based on the survival RuleFit that can construct an interpretable model while maintaining a prediction accuracy similar to that of previous machine learning methods.

To estimate the HTEs using survival RuleFit, we can directly apply survival RuleFit using the meta-algorithms T-learner and S-learner, which provide an HTE estimation framework for existing machine learning methods \cite{kunzel2019}. However, it is important to note that both approaches have drawbacks. First, within the T-learner framework, survival RuleFit models are constructed for the treatment and control groups, and then, HTE is estimated using the difference between the estimates of the treatment and control group models. The primary drawback of this approach is its inability to guarantee that the models for the treatment and control groups share the same base functions. Such differences in base functions could result in differences in the estimates between the treatment and control groups, which are not attributable to the treatment effect. This discrepancy can cause bias in the estimation of the HTE. To address this drawback, Powers et al. (2018)\cite{Powers2018} proposed a new framework called ``shared-basis conditional mean regression’’ by restricting T-learners and creating models with the same base functions in the treatment and control groups. This framework removes bias caused by differences in the model base functions. However, the survival RuleFit was proposed based on the Cox proportional hazard model. Although the shared-basis conditional mean regression framework can restrict the treatment and control group models to sharing the same base functions, differences in the baseline hazard functions may still introduce bias in the HTE estimation. Furthermore, these approaches ignore the possibility of main effect differences between the treatment and control group models, which may also lead to bias in HTE estimation. Danel et al. (2022)\cite{Dandl2022} highlighted the importance of correctly specifying the main effect when estimating HTEs using the Cox proportional hazards model. Second, within the S-learner framework, a survival RuleFit model is constructed using the treatment indicator as a covariate, and then, the HTE is estimated from the difference between the treatment and control group estimates. This approach creates a single model that can automatically model the main and treatment effects. Therefore, the S-learner strategy ensures that the treatment and control group models share the same baseline hazard function and main effects. However, we cannot guarantee that the models of the treatment and control groups have the same base functions. Further details are provided in Section 2.

To address these drawbacks, the proposed method combines the strategies of S-learner and shared-basis conditional mean regression. The S-learner strategy enables the created model to incorporate both main and treatment effects. Consequently, by utilizing the S-learners strategy in the survival RuleFit rule generation algorithm, we can construct a set of rules, including main effect rules, which do not include interactions between treatment indicators and covariates, and treatment effect rules, which do include interactions between treatment indicators and covariates. Similar to the algorithm of survival RuleFit, the main and treatment effect rules were fitted to a sparse linear model to estimate the relationship between the outcomes and the rules. Our proposed method builds a single model consisting of three parts: relationships between the outcome and main effect rules, relationships between the outcome and treatment rules for the control groups, and relationships between the outcome and treatment rules for the treatment groups. To ensure that the model would be interpretable for the HTE and to avoid bias caused by model differences in base functions, we created a model in the shared-basis conditional mean regression framework using group lasso.

The remainder of this paper is organized as follows: Section 2 introduces related work on HTE estimation and illustrates the main purpose of our work. Section 3 explains the original RuleFit method and the proposed method in detail. In Section 4, we describe several simulation studies to compare the prediction performance of the proposed method with those of previous HTE estimation methods. In Section 5, we apply the proposed method to real data and explain how it works. Section 6 summarizes the study and discusses its results.

\section{Previous methods and motivation for proposed method}
In this section, we start by introducing the definition of HTE for survival outcomes. Next, we introduce the RuleFit method for survival outcomes. Finally, we illustrate how the RuleFit method is utilized to estimate HTE for survival outcomes in existing frameworks and discuss their drawbacks. 

\subsection{HTEs for survival data}
To formalize HTEs for survival data, we referred to the definitions used in previous studies\cite{Sugasawa2019,Zhu2020,Cui2023}. Consider a dataset $\{(t_i = \min(t^*_i,c_i),\delta_i,z_i,\bm{x}_i)\}_{i = 1}^N$, where $t_i$ is the observed survival/censoring time for individual $i$, $t^*_i$ is the true survival time for individual $i$, and $c_i$ denotes the censoring time for individual $i$. $\delta_i \in \{0, 1\}$ is the censoring indicator: if individual $i$ experienced an event $\delta_i = 1$, otherwise $\delta_i = 0$. The treatment indicator $z_i \in \{0, 1\}$ assigns individual $i$ to the treatment group if $z_i = 1$ and to the control group if $z_i = 0$. Furthermore, $\bm{x}_i = (x_{i1},x_{i2}, \cdots,x_{ip})^T$ is the $p$-variable covariate vector for individual $i$. The conditional survival function at time $t$ for individual $i$ is defined as:
\begin{align*}
S(t|\bm{x}_i, z_i) := P(T_i>t|\bm{X}_i = \bm{x}_i,Z_i = z_i),
\end{align*}
where $T_i$, $\bm{X}_i$, and $Z_i$ are random variables representing the observed survival/censoring time, covariates, and treatment indicators, respectively. Within the framework of potential outcomes \cite{ROSENBAUM1983}, the HTE for the survival outcome can be defined as the difference in survival rate between the treatment and control group conditions on the covariates $\bm{x}_i$, as follows:
\begin{align}
\Delta(t|\bm{x}_i) &= P(T_i>t|\bm{X}_i = \bm{x}_i,Z_i = 1) - P(T_i>t|\bm{X}_i = \bm{x}_i,Z_i = 0) \notag \\
                   &= S(t|\bm{x}_i,z_i = 1) - S(t|\bm{x}_i,z_i = 0). \label{surv_hte_def}
\end{align}

\subsection{RuleFit for survival outcome} \label{RuleFit_surv}
Given the covariate vector $\bm{x}_i = (x_{i1}, x_{i2}, \cdots, x_{ip})^T \in \mathbb{R}^p$, the survival RuleFit model is defined as: 
\begin{equation}
h_{\mathrm{RuleFit}}(t|\bm{x}_i) =  h_0(t)\exp\left(\sum_{k = 1}^K\theta_{k}r_k(\bm{x}_i) + \sum_{j = 1}^p \theta^*_jl_j(x_{ij})\right), 
\label{surv_hte}
\end{equation}
where $h_0(t)$ is the baseline hazard function, $r_k(\bm{x}_i)$ is the rule term, and $l_j(x_{ij})$ is the linear term. The rule terms $r_k : \mathbb{R}^p \mapsto\mathbb{R}$ represent conjunctions of indicator functions, while the linear terms $l_j : \mathbb{R} \mapsto\mathbb{R}$ are represented by the ``winsorized’’ version. The coefficients $\theta_k \in \mathbb{R} (k = 0, 1, \cdots, K)$ and $\theta^*_j \in \mathbb{R} (j = 1, \cdots, p)$ correspond to the rule term coefficients and linear term coefficients, respectively. The rule terms $r_k$ and linear terms $l_j$ are explained in detail below.

\paragraph{Rule terms} The $k$-th rule term is defined as:
\begin{align*}
r_k(\bm{x}_i) = \prod_{j=1}^pI(x_{ij} \in S_{jk}),
\end{align*}
where $I(\cdot)$ is an indicator function that returns 1 if the condition within parentheses is true and 0 if it is false. $S_j$ is the set of all possible values for $x_j$ and the subset $S_{jk}\subset S_j$ is defined by the interval:
\begin{align*}
S_{jk} = [x_{jk}^-, x_{jk}^+),
\end{align*}
where $x_{jk}^-$ and $x_{jk}^+$ represent the lower and upper bounds of $x_j$ defined by the $k$th rule term. 

\paragraph{Linear terms} Friedman and Popescu (2008)\cite{Friedman2008} noted that including the basis function with different structures can increase the prediction accuracy, and various linear terms were chosen for addition to the RuleFit model. However, directly adding linear terms can reduce the model’s robustness against outliers compared to a model composed solely of rule terms. To address this issue, the linear terms were given a ``winsorized’’ form as:
\begin{align}
l_j(x_{ij}) = \min(\delta^+_j,\max(\delta_j^-,x_{ij})),
\end{align}
where $\delta_j^+$ and $\delta_j^-$ are the thresholds for determining the outliers defined by the $q$-quantile and $(1-q)$-quantile of $x_j$, respectively, with a recommended value of $q = 0.025$. To give the rule and linear terms an equal chance of being selected, the linear term was normalized as follows:
\begin{align*}
l_j(x_{ij}) \leftarrow 0.4\cdot l_j(x_{ij})/std(l_j(x_{ij})),
\end{align*}
where $std(\cdot)$ represents the standard deviation and 0.4 is the average standard deviation of the rule terms under the assumption that the support of the rule terms $r_k(\bm{x}_i)$ from the training data
\begin{align}
    \rho_k = \frac{1}{N} \sum_{i = 1}^N r_k(\bm{x}_i) \label{support}
\end{align}
are distributed uniformly from $U(0,1)$.

\subsection{Drawbacks of HTE estimation using the existing survival RuleFit model} 
According to the definition of the survival HTE in Eq. \ref{surv_hte_def}, the survival HTE can be derived from the difference in the conditional survival function estimates at time $t$ between the treatment and control groups. Therefore, we can use the survival RuleFit model for the treatment and control groups and calculate the HTE from the difference between the estimates for the treatment and control groups. There are two common frameworks used for addressing such tasks: T-learner and S-learners in K\"{u}nzel et al. (2019)\cite{kunzel2019}. Both provide a simple framework for easily extending existing regression approaches to HTE estimation. Next, we briefly introduce the estimation of HTE using survival RuleFit based on T-learner and S-learner, and discuss their drawbacks for our purposes.

\paragraph{T-learner for survival RuleFit} Two survival RuleFit models were fitted to the treatment and control groups. The conditional mean survival function for the control group can be estimated as:
\begin{align*}
S^{(0)}(t|\bm{x}_i) = \exp\left(-\int_0^t h^{(0)}_{\mathrm{RuleFit}}(u|\bm{x}_i)du\right) =\exp\left(-\int_0^th_0^{(0)}(u)\exp\left(\sum_{k = 1}^K\hat{\gamma}^{(0)}_{k}r_k^{(0)}(\bm{x}_i) + \sum_{j = 1}^p \hat{\gamma}^{*(0)}_jl_j(x_{ij})\right)du\right),
\end{align*}
and that for the treatment group can be estimated as:
\begin{align*}
S^{(1)}(t|\bm{x}_i) = \exp\left(-\int_0^t h^{(1)}_{\mathrm{RuleFit}}(u|\bm{x}_i)du\right) =\exp\left(-\int_0^th_0^{(1)}(u)\exp\left(\sum_{k = 1}^K\hat{\gamma}^{(1)}_{k}r_k^{(1)}(\bm{x}_i) + \sum_{j = 1}^p \hat{\gamma}^{*(1)}_jl_j(x_{ij})\right)du\right),
\end{align*}
where $h_0^{(0)}(t)$ and $h_0^{(1)}(t)$ are the baseline hazard functions for the treatment and control groups, respectively; $r_k^{(0)}(\bm{x}_i)$ and $r_k^{(1)}(\bm{x}_i)$ are the rule terms for the treatment and control groups, respectively; $\hat{\gamma}_k^{(0)}$ and $\hat{\gamma}_k^{(1)}$ are the estimated coefficients of the rule terms for the treatment and control groups; and $\hat{\gamma}_k^{*(0)}$ and $\hat{\gamma}_k^{*(1)}$ are the estimated coefficients of the linear terms for the treatment and control groups. Then, the HTE can be estimated as:
\begin{align*}
\hat{\Delta}(t|\bm{x}_i) = S^{(1)}(t|\bm{x}_i) - S^{(0)}(t|\bm{x}_i).
\end{align*}
The main drawback of this approach is that the fitted models for the treatment and control groups can consist of different base functions. Powers et al. (2018)\cite{Powers2018} pointed out that such a difference could lead to a difference in the treatment and control group estimates not solely attributable to the effect of the treatment, and this discrepancy can cause bias in the estimation of HTE. The proposed framework, based on the T-learner framework and named the ``shared-basis conditional mean regression,’’ addresses this issue by adding a restriction that ensures the treatment and control group models share the same base function. This approach reduces the bias arising from differences in the base functions of the models between the treatment and control groups. It also ensures the comparability of rule terms between the treatment and control group models and allows the estimation of the HTE for each rule. Therefore, by applying the survival RuleFit within the shared-basis conditional mean regression framework, the estimated HTE becomes interpretable contingent on its base function. However, both T-learners and shared-basis conditional mean regression construct separate models for the treatment and control groups without constraining the main effects to be the same. This may lead to inaccuracies in HTE estimation owing to differences in the main effects.

\paragraph{S-learner for survival RuleFit} The survival RuleFit model was fitted to the data, with the treatment indicator considered as a covariate. Subsequently, the survival RuleFit model, including the interactions between treatment and covariates, was created. The estimates of the conditional mean survival function were denoted as:
\begin{align*}
S(t|\bm{x}_i, z_i) = \exp\left(-\int_0^t h_{\mathrm{RuleFit}}(u|\bm{x}_i, z_i)du\right) =\exp\left(-\int_0^th_0(u)\exp\left(\sum_{k = 1}^K\hat{\gamma}_{k}r_k(\bm{x}_i, z_i) + \sum_{j = 1}^p \hat{\gamma}^*_jl_j(x_{ij})\right)du\right),
\end{align*}
and the HTE was estimated as: 
\begin{align*}
\hat{\Delta}(t|\bm{x}_i) = S(t|\bm{x}_i,z_i = 1) - S(t|\bm{x}_i, z_i = 0).
\end{align*}
Compared to the T-learner framework, S-learner creates a single model consisting of the interaction between the treatment and covariates. Therefore, the S-learner strategy ensures that the treatment and control groups have similar main effects. However, the survival RuleFit creates rules using tree-based learners, and the interaction rules between the treatment and covariates take the forms of ``$I(z = 1)I(x < c)$’’ or ``$I(z = 0)I(x < c)$.’’ Consequently, for the subgroups where $x < c$, we can only obtain the effect in the treatment group ($z = 1$) or in the control group ($z = 0$) at once and the other part will be ignored. Therefore, the HTE for the subgroups ``$x < c$’’ cannot be estimated, making the relationships between $x < c$ and HTE uninterpretable. Thus, the application of survival RuleFit in the S-learner framework cannot be used to build an interpretable model for HTE and loses the primary advantage of survival RuleFit.

\section{Proposed method}
In the previous section, we discussed the drawbacks of applying survival RuleFit for HTE estimation. First, within the framework of a shared-basis conditional mean regression, an interpretable HTE estimation model can be constructed. However, this approach may still suffer from potential inaccuracies owing to differences in the main effects between the models for the treatment and control groups. Second, within the framework of S-learner, the models created for the treatment and control groups shared the same main effects, avoiding inaccuracies arising from differences in main effects. However, the estimated HTE cannot be interpreted using this approach. To address these issues, we combine these approaches and propose a novel method for survival HTE estimation. In this section, we first introduce our proposed model and then explain how to estimate the parameters.

\subsection{Model of the proposed method}
We propose a method for survival HTE estimation based on the survival RuleFit model. Unlike the survival RuleFit model shown in Eq. \ref{surv_hte}, our proposed method models the main effect, the effect of the treatment groups, and the effect of the control groups simultaneously. Given the dataset $\{(t_i,\delta_i,z_i,\bm{x}i)\}_{i=1}^N$, the model of the proposed method is as follows:
\begin{align}
h(t|\bm{x}_i, z_i) &= h_0(t)\exp\Biggl[\sum_{k = 1}^{K^{\dag}}\theta_kr^{\dag}_k(\bm{x}_i) + \sum_{j = 1}^p\theta^*_j l_j(x_{ij}) \notag\\
               &+ I(z_i = 1)\biggl\{\sum_{k = 1}^{K^{\ddag}}\alpha_k r_k^{\ddag} (\bm{x}_i) + \sum_{j = 1}^p \alpha^*_j l_j(x_{ij})\biggr\} \notag\\
               &+ I(z_i = 0)\biggl\{\sum_{k = 1}^{K^{\ddag}} \beta_k r_k^{\ddag} (\bm{x}_i) + \sum_{j = 1}^p \beta^*_j l_j(x_{ij})\biggr\}\Biggr], \label{prop}
\end{align}
where $r_k^{\dag}(\bm{x}_i)$ is the rule term for the main effect and $r_k^{\ddag}(\bm{x}_i)$ is the rule term for the treatment effect. $\theta_k, \alpha_k, \beta_k \in \mathbb{R} (l = 1,2,\cdots, L; k = 1,2, \cdots, K)$ and $\theta^*_j, \alpha^*_j, \beta^*_j$ are the coefficients of the rule terms and linear terms, respectively. Furthermore, to ensure that the models for the treatment and control groups had the same structure and to maintain model interpretability, our proposed method was created under the shared-basis conditional mean regression framework. Therefore, a constraint on the coefficients for the rule terms,
\begin{align*}
\begin{cases}
\hat{\alpha}_k = 0 \land \hat{\beta}_k =0\\
\hat{\alpha}_k \neq 0 \land \hat{\beta}_k \neq 0
\end{cases},
\end{align*}
and a constraint on the coefficients for the linear terms,
\begin{align*}
\begin{cases}
\hat{\alpha}^*_k = 0 \land \hat{\beta}^*_k =0\\
\hat{\alpha}^*_k \neq 0 \land \hat{\beta}^*_k \neq 0
\end{cases},
\end{align*}
were added to the model. Therefore, the estimates of the conditional hazard function for the control groups can be estimated as: 
\begin{align*}
\hat{h}(t|\bm{x}_i, z_i = 0) &=  h_0(t) \exp\left\{\left(\sum_{k = 1}^{K^{\dag}}\hat{\theta}_{k}r_k^{\dag}(\bm{x}_i) + \sum_{j = 1}^p \hat{\theta}^*_jl_j(x_{ij})\right) + \left(\sum_{k = 1}^{K^{\ddag}}\hat{\beta}_{k}r_k^{\ddag}(\bm{x}_i) + \sum_{j = 1}^p \hat{\beta}^*_jl_j(x_{ij})\right)\right\}.
\end{align*}
The estimates of the conditional hazard function for the treatment groups can be expressed as 
\begin{align*}
\hat{h}(t|\bm{x}_i, z_i = 1) &=  h_0(t) \exp\left\{\left(\sum_{k = 1}^{K^{\dag}}\hat{\theta}_{k}r_k^{\dag}(\bm{x}_i) + \sum_{j = 1}^p \hat{\theta}^*_jl_j(x_{ij})\right) + \left(\sum_{k = 1}^{K^{\ddag}}\hat{\alpha}_{k}r_k^{\ddag}(\bm{x}_i) + \sum_{j = 1}^p \hat{\alpha}^*_jl_j(x_{ij})\right)\right\}.
\end{align*}
Therefore, the HTE can be estimated as: 
\begin{align}
\hat{\Delta}(t|\bm{x}_i) &= S(t|\bm{x}_i, z_i = 1) - S(t|\bm{x}_i, z_i = 0) \notag \\
                         &= \exp\left(-\int_0^t \hat{h}(u|\bm{x}_i, z_i = 1) du\right) -  \exp\left(-\int_0^t \hat{h}(u|\bm{x}_i, z_i = 0) du\right). \label{surv_hte_prop}
\end{align}

Furthermore, the proposed method (Eq.\ref{prop}) is also based on the Cox proportional hazard model. Thus, the difference in survival rates between the treatment and control groups could also be interpreted using the hazard ratio, and the HTE estimates can be interpreted based on the estimated hazard ratio as
\begin{align*}
\frac{\hat{h}(t|\bm{x}_i, z_i = 1)}{\hat{h}(t|\bm{x}_i, z_i = 0)} &= \frac{h_0(t) \exp\left\{\left(\sum_{k = 1}^{K^{\dag}}\hat{\theta}_{k}r_k^{\dag}(\bm{x}_i) + \sum_{j = 1}^p \hat{\theta}^*_jl_j(x_{ij})\right) + \left(\sum_{k = 1}^{K^{\ddag}}\hat{\alpha}_{k}r_k^{\ddag}(\bm{x}_i) + \sum_{j = 1}^p \hat{\alpha}^*_jl_j(x_{ij})\right)\right\}}{h_0(t) \exp\left\{\left(\sum_{k = 1}^{K^{\dag}}\hat{\theta}_{k}r_k^{\dag}(\bm{x}_i) + \sum_{j = 1}^p \hat{\theta}^*_jl_j(x_{ij})\right) + \left(\sum_{k = 1}^{K^{\ddag}}\hat{\beta}_{k}r_k^{\ddag}(\bm{x}_i) + \sum_{j = 1}^p \hat{\beta}^*_jl_j(x_{ij})\right)\right\}} \notag \\
&= \exp\left\{\left(\sum_{k = 1}^{K^{\ddag}}\hat{\alpha}_{k}r_k^{\ddag}(\bm{x}_i) + \sum_{j = 1}^p \hat{\alpha}^*_jl_j(x_{ij})\right) - \left(\sum_{k = 1}^{K^{\ddag}}\hat{\beta}_{k}r_k^{\ddag}(\bm{x}_i) + \sum_{j = 1}^p \hat{\beta}^*_jl_j(x_{ij})\right)\right\} \notag \\
&= \exp\left\{\sum_{k = 1}^{K^{\ddag}}\left(\hat{\alpha}_k - \hat{\beta}_{k}\right)r_k^{\ddag}(\bm{x}_i) + \sum_{j = 1}^p \left(\hat{\alpha}^*_j - \hat{\beta}^*_j\right)l_j(x_{ij})\right\}.
\end{align*}
Consequently, the base functions $\{r_k^{\ddag}(\bm{x}_i)\}_{k=1}^{K^{\ddag}}$ and $\{l_j(x_{ij})\}_{j=1}^p$ and corresponding differences in the coefficients $\{(\hat{\alpha}_k - \hat{\beta}_k)\}_{k=1}^{K^{\ddag}}$ and $\{(\hat{\alpha}^*_j - \hat{\beta}^*_j)\}_{j=1}^p$ can be used to interpreted the estimated HTE. The next section provides a detailed explanation of the parameter estimation process.

\subsection{Algorithm of proposed method}
To construct the proposed model, we developed an algorithm based on the survival RuleFit algorithm. The proposed algorithm consists of three steps: 1) rule generation, 2) rule division, 3) and rule ensemble. We provide a brief overview of our proposed algorithm and illustrate its differences from survival RuleFit. A detailed description of this algorithm is provided in the following subsections.

\noindent{\bf{Step1: Rule generation}}: The purpose of this step is to generate candidate rule terms for the proposed method. To ensure the proposed model's interpretation of survival HTE, these rules should be related to treatment effects. In this step, we followed the same algorithms as used in survival RuleFit, but we considered the treatment indicators as covariates during model construction. In this manner, we automatically obtained rules related to the treatment indicators that are considered to be related to survival HTE.

\noindent{\bf{Step2: Rule division}}: The purpose of this step is to divide the set of rule terms generated in the rule generation step into two subsets: one for the main effect and the other for the treatment effect. This step is the main difference between survival RuleFit and the proposed method. The resulting sets of main effect rules and treatment effect rules were then used to model the main and treatment effects, respectively, in the subsequent steps.

\noindent{\bf{Step3: Rule ensemble}}: The purpose of this step is to select base functions and estimate coefficients for the final model. In survival RuleFit, this is achieved by fitting the rules and linear terms to sparse linear models using lasso. This simplifies the model, improves interpretability, and prevents overfitting. However, unlike the survival RuleFit, the proposed method requires the selection of treatment effect rules for both the treatment and control groups to ensure that their models have the same rule terms to maintain the interpretability of the estimated HTE. This was achieved by grouping the treatment effect rules for both groups and using a group lasso to select the base function and coefficient estimation.

\subsubsection{Rule generation}
In this step, candidate rules for the proposed model are constructed. Given the dataset $\{(t_i, \delta_i, z_i, \bm{x}_i)\}_{i = 1}^N$, generalized boosted models (GBMs)\cite{Ridgeway2007} are fitted to the dataset to construct rules related to the treatment effect. The treatment indicator $z_i$ was regarded as a covariate in model building. The GBM model $F_M(\bm{x}_i,z_i)$ with $M$ tree base learner is built in the form of $F_M(\bm{x}_i,z_i) = \sum_{m = 1}^M f_m(\bm{x}_i, z_i)$, where $M$ is the number of the base learner; the $m$-th base learner $f_m(\bm{x}_i,z_i) = \sum_{d = 1}^{D_m} \gamma_d I((\bm{x}_i,z_i)\in R_d)$ is the decision tree, consisting of disjointed partitioned regions $R_{1},R_{2},\cdots, R_{D_m}$; $\gamma_{d}$ is the weight of the corresponding region $R_{d}$; and $D_m$ is the number of regions. Therefore, according to these steps, we obtained a set of tree-based learners $\{f_m(\bm{x}_i,z_i)\}_{m=1}^M$. Then, to create the candidate rules for the proposed method, the whole tree base learners $\{f_m(\bm{x}_i,z_i)\}_{m=1}^M$ are decomposed into a set of rules ${r_k(\bm{x}_i, z_i)}_{k = 1}^K$ with the number of rules $K = \sum_{m=1}^M 2(D_m -1)$, in same manner as used in the survival RuleFit \cite{Fokkema2020}. 

\begin{algorithm}[tb]
\caption{Rule generation}
\label{Rulefit_create}
\begin{algorithmic}[1]
\STATE \textbf{Input} : Dataset $\{t_i, \delta_i, z_i, \bm{x}_i\}_{i = 1}^N$, number of tree base learners $M$, mean depth of tree base learners $\bar{L}$, shrinkage rate $v$, and the training sample fraction for each tree-based learner $\eta$
\STATE Initialize the model $F_0(\bm{x}_i,z_i) = 0$
\FOR{$m = 1$ to $M$}
\STATE For $i = 1, 2, \cdots, N$, compute the gradient for $F_{m-1}(\bm{x}_i,z_i)$ as:
\begin{align*}
r^{(m)}_i = \delta_i - \sum_{i^* = 1}^N \delta_{i^*} \frac{I(t_i > t_{i^*})\mathrm{exp}(F_{m-1}(\bm{x}_i,z_i))}{\sum_{i^{**}=1}^NI(t_{i^{**}} > t_{i^*})\mathrm{exp}(F_{m-1}(\bm{x}_{i^{**}},z_{i^{**}}))} 
\end{align*}
\STATE Determine the number of terminal nodes for the tree-based learner, $D_m = 2 + \mathrm{floor}(u)$, \\where $u \sim \mathrm{exponential}(1/(\bar{L} - 2))$
\STATE Fit a regression tree $f_{m-1}(\bm{x}_i,z_i)$ to the gradient $r^{(m)}$, giving the terminal regions $R_d, d = 1, 2, \cdots D_m$
\STATE For $d = 1,2,\cdots,D_m$, estimate the value of region $R_d$ as 
\begin{align*}
\hat{\gamma}_d = \argmin_{\gamma_d}\sum_{(\bm{x}_i,z_i) \in R_d}\sum_{i \in H} (r^{(m)}_i - \gamma_d)^2,
\end{align*}
where $H \subset \{1,2,\cdots, N\}$ is a sample-set randomly drawn from the data and $|H| = \lfloor \eta N \rfloor$
\STATE Update $F_{m}(\bm{x}_i,z_i) = F_{m-1}(\bm{x}_i,z_i) + v\cdot \sum_{d=1}^{D_m}\hat{\gamma}_d I((\bm{x}_i,z_i) \in R_d) = F_{m-1}(\bm{x}_i,z_i) + vf_m(\bm{x}_i,z_i)$
\ENDFOR
\STATE For $m = 1,2,\cdots, M$, traverse the tree $f_m(\bm{x}_i, z_i)$ to create the rule set $\{r_k(\bm{x}_i, z_i)\}_{k=1}^{K^{(m)}}$, where $K^{(m)} = 2(D_m - 1)$ 
\STATE Aggregate every rule set $\{r_k(\bm{x}_i, z_i)\}_{k=1}^{K^{(1)}}, \{r_k(\bm{x}_i, z_i)\}_{k=1}^{K^{(2)}},\cdots ,$ and $\{r_k(\bm{x}_i, z_i)\}_{k=1}^{K^{(M)}}$ into $\{r_{k}(\bm{x}, z_i)\}_{k=1}^{K}$, where $K = \sum_{m =1}^M K^{(m)}$.
\STATE \textbf{Output} $\{r_{k}(\bm{x}_i,z_i)\}_{k=1}^{K}$
\end{algorithmic}
\end{algorithm}

\begin{figure}[t]
 \centering
 \includegraphics[width = \linewidth]{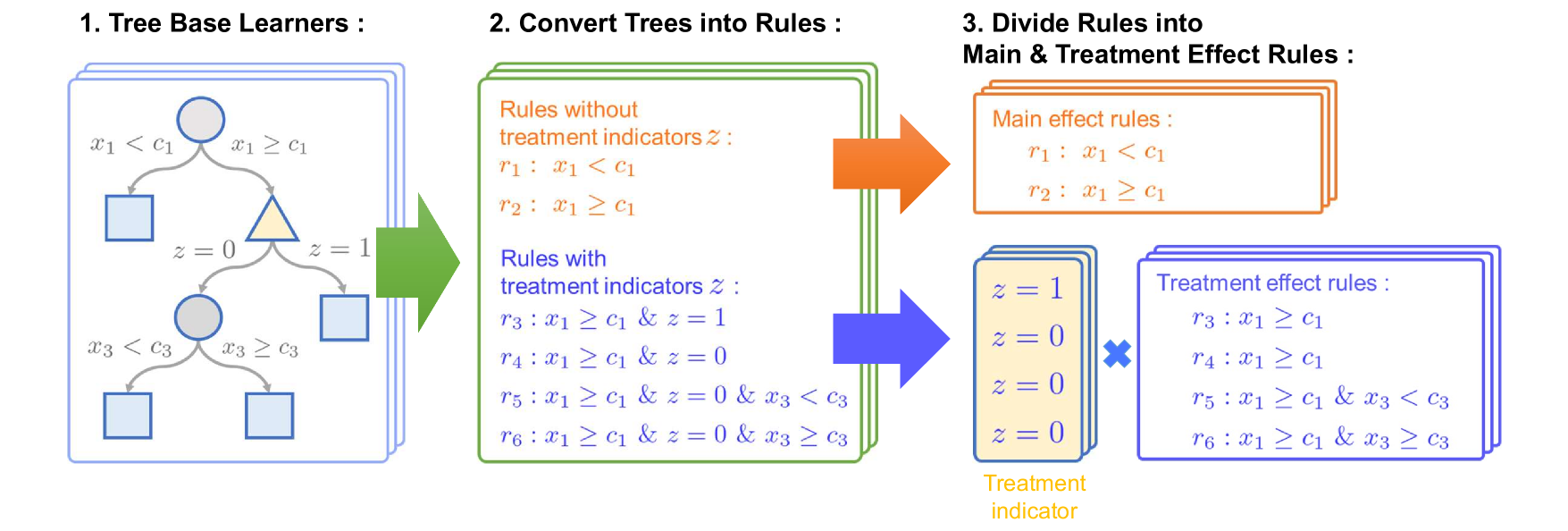}
 \caption{Example illustrating tree decomposition and rule division for the proposed method}
 \label{fig1}
\end{figure}

The algorithm for generating rules is presented in Algorithm \ref{Rulefit_create} and is explained in detail here. The lines in this paragraph correspond to those in Algorithm \ref{Rulefit_create}. Line 1 specifies the input for rule generation, including the dataset and several hyperparameters for the GBM. From lines 3 to 9, we update the parameters in a manner similar to that of the GBM algorithm. Line 5 randomly determines the depth of the base learner. This process is crucial to survival RuleFit and makes it possible to obtain rules with various depths. In the common GBM algorithm, the depth of each base learner is small and fixed. Therefore, the maximum depth of the created rules is also small, and a linear combination of these rules cannot capture high-order interactions. Thus, in the survival RuleFit algorithm, the depth of each base learner is randomly determined. However, to avoid affecting the performance of the GBM, the depth of most base learners remains small, and only a few of them have a large depth. To achieve this, the depth of each base learner is randomly drawn from the exponential distribution $\mathrm{exponetial}(1/(\bar{L} - 2))$, where $\mathrm{floor}(u)$ represents the largest integer less than or equal to $u$. From lines 10 to 12, we decompose the tree-based learners into rules, as shown in Fig\ref{fig1} and output the candidate rules set. 

\subsubsection{Rule division}
In this step, the created rules $\{r_k(\bm{x}_i,z_i)\}_{k=1}^K$ are divided into rules for the main effect $\{r_k^{\dag}(\bm{x}_i)\}_{k=1}^{K^\dag}$ and treatment effect $\{r_k^{\ddag}(\bm{x}_i)\}_{k=1}^{K^\ddag}$. In the rule generation step, a GBM was created with the treatment indicator as a covariate. Consequently, tree-based learners of the GBM model can automatically detect the interaction effect between the treatment and other covariates, allowing us to divide the rules into main and treatment effect rules. We present a simple example to briefly illustrate this process in Fig. \ref{fig1}.

The $k$th rule in the rule set $\{r_k(\bm{x}_i,z_i)\}_{k=1}^K$ can be expressed as
\begin{align*}
r_k(\bm{x}_i,z_i) =
\begin{cases}
&I(z_i \in S_k^{(z)})r_k^{\dag}(\bm{x}_i), \qquad \mathrm{if} \ S_k^{(z)} = \{0,1\} \\
&I(z_i \in S_k^{(z)})r_k^{\ddag}(\bm{x}_i),\qquad \mathrm{if} \ S_k^{(z)} = \{0\} \ \mathrm{or} \ S_k^{(z)} = \{1\}
\end{cases},
\end{align*}
where $S_k^{(z)}$ is the set of available values of the treatment indicator. Therefore, there is an interaction effect between the treatment indicator and $r_k^{\ddag}(\bm{x}_i)$, but no interaction effect between the treatment indicator and $r_k^{\dag}(\bm{x}_i)$. In this sense, we define $r_k^{\ddag}(\bm{x}_i)$ as the treatment effect rule and $r_k^{\dag}(\bm{x}_i)$ as the main effect rule. Therefore, in this step, we finally produce the set of rules for the treatment effect $\{r_k^{\ddag}(\bm{x}_i)\}_{k=1}^{K^{\ddag}}$ and that for the main effect $\{r_k^{\dag}(\bm{x}_i)\}_{k=1}^{K^{\dag}}$, where $K = K^{\ddag} + K^{\dag}$.

\begin{algorithm}[tb]
\caption{Rule division}
\label{Rulefit_divide}
\begin{algorithmic}[1]
\STATE \textbf{Input} : Candidate rules set $\{r_{k}(\bm{x}_i,z_i)\}_{k=1}^{K}$.
\STATE \textbf{Initialize} : $R^{(main)} \leftarrow \phi $, $R^{(treat)} \leftarrow \phi$
\FOR{$k = 1$ to $K$}
\IF{$S_k^{(z)} = \{0,1\}$}
\STATE $r_k(\bm{x}_i,z_i) = I(z_i \in S_k^{(z)})r_k^{\dag}(\bm{x}_i)$
\STATE $R^{(main)} \leftarrow R^{(main)} \cup \{r_k^{\dag}\}$
\ELSIF{$S_k^{(z)} = \{0\} \ \mathrm{or} \ S_k^{(z)} = \{1\}$}
\STATE $r_k(\bm{x}_i,z_i) = I(z_i \in S_k^{(z)})r_k^{\ddag}(\bm{x}_i)$
\STATE $R^{(treat)} \leftarrow R^{(treat)} \cup \{r_k^{\ddag}\}$
\ENDIF
\ENDFOR
\STATE \textbf{Output} : $R^{(main)}$ and $R^{(treat)}$, where $|R^{(main)}| = K^{\dag}$ and $|R^{(treat)}| = K^{\ddag}$ 
\end{algorithmic}
\end{algorithm}

The algorithm for rule division is presented in Algorithm \ref{Rulefit_divide} and is explained in detail here. The lines in this paragraph correspond to those in Algorithm \ref{Rulefit_divide}. Line 1 specifies the input for the rule-division process, which is a set of rules created during the rule-generation process. From lines 3 to 10, we divided the rules into a set of main effect rules and a set of treatment effect rules based on whether any interactions involved the treatment indicators.

\subsubsection{Rule ensemble}
In this step, the base functions (rule and linear terms) for the proposed method were selected, and the coefficients for the proposed method were estimated. First, we added the ``winsorized’’ versions of the linear terms as base functions in the same manner as that used in the survival RuleFit. We then selected the optimal model and estimated the coefficients simultaneously using a group lasso. To illustrate how the group lasso works in the proposed method, we first introduce the group lasso.

\paragraph{Group lasso} Group lasso \cite{Yuan2006} is a regularized method which allows the variable selection of a predefined group of variables in regression models. Given the data $\{y_i,\bm{g}_{i1},\bm{g}_{i2},\cdots, \bm{g}_{iL}\}_{i=1}^N$ consisting of $L$ groups of variables, where the grouped variable $\bm{g}_{\ell}\in \mathbb{R}^{p_{\ell}} ({\ell} = 1,2,\cdots,L)$ and $p_{\ell}$ is the number of variables in group ${\ell}$, the estimated intercept $\theta_0 \in \mathbb{R}$ and coefficients of the grouped variable $\bm{\theta}_{\ell}\in\mathbb{R}^{p_{\ell}} ({\ell}=1,2,\cdots,J)$ are then defined as:

\begin{align*}
(\hat{\theta}_0,\{\hat{\bm{\theta}}_{\ell}\}_{{\ell}=1}^L) = \argmin_{\theta_0,\{\bm{\theta}_{\ell}\}_{{\ell}=1}^L} \mathcal{L}(\theta_0,\{\bm{\theta}_{\ell}\}_{{\ell}=1}^L) + \sum_{{\ell} = 1}^L\lambda \sqrt{p_{\ell}}||\bm{\theta}_{\ell}||_2,
\end{align*}
where $||\bm{\theta}_{\ell}||_2$ is the Euclidean norm of the coefficients, $\bm{\theta}_{\ell}$ and $\lambda \in \mathbb{R}^+$ are the tuning parameters, and $\mathcal{L}(\theta_0,\bm{\theta}_{\ell})$ is the loss function; for example, 
\begin{align*}
\mathcal{L}(\theta_0,\{\bm{\theta}_{\ell}\}_{{\ell}=1}^L) = \frac{1}{2}\sum_{i=1}^N (y_i - \theta_0 - \sum_{{\ell} = 1}^L {\bm{g}_{i{\ell}}}^T\bm{\theta}_{\ell})^2
\end{align*}
in a linear regression model.

\paragraph{Application of the proposed model}
Here, we apply the group lasso to estimate the coefficients for the main effect $\theta_k$ and $\theta^*_j$ for the treatment groups $\alpha_k$ and $\alpha^*_j$, and for the control groups $\beta_k$ and $\beta^*_j$, respectively. As mentioned above, to maintain the interpretability of the proposed model, we grouped the treatment effect rules for the treatment group ($z_i = 1$) and control group ($z_i = 0$) as:
\begin{align*}
\bm{u}_{ik} & = (z_ir_k^{\ddag}(\bm{x}_i), (1-z_i)r_k^{\ddag}(\bm{x}_i))^T \quad, k= 1, 2, \cdots, K^{\ddag} \ \mathrm{and}\\
\bm{u}_{ij}^* & = (z_il_j(x_{ij}), (1-z_i)l_j(x_{ij}))^T \quad,j = 1,2, \cdots, p;\ i = 1, 2, \cdots,N,
\end{align*}
respectively. Additionally, we denote the coefficients of the grouped treatment effect rules as:
\begin{align*}
\bm{\theta}_{k} & = (\alpha_k, \beta_k)^T \quad, k= 1, 2, \cdots, K^{\ddag} \ \mathrm{and}\\
\bm{\theta}_{j}^{**} & = (\alpha^*_j, \beta^*_j)^T \quad,j = 1,2, \cdots, p. 
\end{align*}

The proposed model (Eq. \ref{prop}) can be rewritten as:
\begin{align*}
h(t|\bm{x}_i, z_i) &= h_0(t)\exp\Biggl[\sum_{k = 1}^{K^{\dag}}\theta_kr^{\dag}_k(\bm{x}_i) + \sum_{j = 1}^p\theta^*_j l_j(x_{ij}) + \biggl\{\sum_{k = 1}^{K^{\ddag}}{\bm{\theta}_{k}}^T\bm{u}_{ik}  + \sum_{j = 1}^p{\bm{\theta}_{j}^{**}}^T\bm{u}^*_{ij}\biggr\}\Biggr] %\label{propmodel2},
\end{align*}
and using the adaptive group lasso, the coefficients are estimated as
\begin{align}
&\left(\{{\hat{\theta}}_k\}_{k=1}^{K^{\dag}}, \{{\hat{\theta}}_j^*\}_{j=1}^p, \{{\hat{\bm{\theta}}}_k\}_{k=1}^{K^{\ddag}}, \{{\hat{\bm{\theta}}}_j^{**}\}_{j=1}^p\right) \\
&= \argmin_{\{{\theta}_k\}_{k=1}^{K^{\dag}}, \{{\theta}_j^*\}_{j=1}^p, \{{\bm{\theta}}_k\}_{k=1}^{K^{\ddag}}, \{{\bm{\theta}}_j^{**}\}_{j=1}^p} \frac{2}{N}\bigg\{\sum_{i=1}^N\delta_i\biggl(\sum_{k = 1}^{K^{\dag}}\theta_{k}r_k(\bm{x}_i) + \sum_{j = 1}^p\theta_{j}^*l_j(x_{ij}) + \sum_{k = 1}^{K^{\ddag}}{\bm{\theta}_{k}}^T\bm{u}_{ik}  + \sum_{j = 1}^p{\bm{\theta}_{j}^{**}}^T\bm{u}^*_{ij} \biggr) \notag \\
&- \sum_{i = 1}^N\delta_i\log\biggl(\sum_{m\in R_{(i)}}\exp\biggl(\sum_{k = 1}^{K^{\dag}}\theta_{k}r_k(\bm{x}_m) + \sum_{j = 1}^p\theta_{j}^*l_j(x_{mj}) + \sum_{k = 1}^{K^{\ddag}}{\bm{\theta}_{k}}^T\bm{u}_{mk}  + \sum_{j = 1}^p{\bm{\theta}_{j}^{**}}^T\bm{u}^*_{mj} \biggr)\biggr)\bigg\} \notag \\
&+\lambda\left(\sum_{k = 1}^{K^{\dag}}||\theta_k||_2 + \sum_{j = 1}^p||\theta_j^*||_2 + \sum_{k =1}^K\sqrt{2}||\bm{\theta}_{k}||_2 + \sum_{j =1}^p\sqrt{2}||\bm{\theta}^{**}_{j}||_2\right), \label{adg}
\end{align}
where $R_{(i)}$ is the set of at-risk subjects at time $t_i$. The HTE can then be estimated using Eq. \ref{surv_hte_prop} based on the estimated coefficients.

\subsection{Interpretation Tools}
To allow the results of RuleFit to be more easily understood, Friedman and Popescu (2008)\cite{Friedman2008} provided several interpretation tools, including base function importance and variable importance. The importance of the base function was used to evaluate the contribution of each base function to the outcome, whereas the variable importance was used for each variable. In this study, we modified the original interpretation tools and applied them to help interpret the application results. We introduce these concepts in detail below.

\subparagraph{Base function importance} The based function importance includes the importance of the rule terms and linear terms. A high base function importance value indicates that the corresponding base function is closely related to the HTE, whereas a low value indicates that it contributes little to the HTE. Here, we modified the base function importance based on the original functions and provided the importance of the rule and linear terms for our proposed method as follows: 
\begin{align}
I_k &= |\hat{\alpha}_k -\hat{\beta}_k|\cdot\sqrt{\varrho_k(1-\varrho_k)} \quad \mathrm{and} \label{rule_imp} \\
I_j &= |\hat{\alpha}^*_j - \hat{\beta}^*_j|\cdot|l_j(x_j)-\bar{l}_j|, \label{lin_imp}
\end{align}
respectively, where $\varrho_k$ is the support for the rules, as shown in Eq. \ref{support}.

\subparagraph{Variable importance} Variable importance is a general approach that helps to interpret black-box machine learning methods, and it can provide information on variables that are most strongly related to the outcomes. Although the proposed method is interpretable, using the variable importance makes our results easier to understand. Thus, we referred to the variable importance of the original RuleFit and define it as follows:
\begin{align}
I^*_{j}(\bm{x}) = I_j(\bm{x}) + \sum_{x_j\in r_k} \frac{I_k(\bm{x})}{m_k}, \label{var_imp}
\end{align}
where the first term $I_j(\bm{x})$ is the importance of the $j$ th linear term, and the second term is the sum of the importance of the rules that contain $x_j (x_j\in r_k)$, with each rule importance divided by the total number of variables $x_j$ used to define the rule.

\section{Simulation study}
Several artificial datasets were used to evaluate the performance of the proposed method under various conditions. 
\begin{comment}
We also compared the performance of the proposed method with those of several existing methods based on Cox hazard models, including virtual twins\cite{Foster2011}, enriched random survival forests, random survival forests using S-learners \cite{kunzel2019}, and causal survival forests\cite{Cui2023}. 
\end{comment}
We provide a detailed description of the design of the artificial data for each simulation and present the application results of each method for each artificial dataset.

\subsection{Simulation design}
For each simulation study, the dataset $\{(t_i,\delta_i,z_i,\bm{x}_i)\}_{i=1}^N$ with $N = 1000$ samples, was generated for the training and test datasets. For each individual $i$, $t_i$ is the observed survival time, $\delta_i$ is the censoring indicator, $z_i$ is the treatment indicator, and $\bm{x}_i = (x_{i1},x_{i2},\cdots, x_{ip})$ is a vector of explanatory variables. The number of covariates was set to $p = 15$. The simulation design is described in detail below.

\paragraph{Explanatory variables}
The explanatory variables $\bm{x}_i = (x_{i1},x_{i2},\cdots, x_{ip})$ consist of both continuous and binary values, where the odd-numbered covariates have continuous values distributed according to the standard normal distribution $N(0, 1)$ and the even-numbered covariates have binary values distributed according to the Bernoulli distribution $B(0.5)$. 

\paragraph{Treatment indicator}
The treatment indicator $t_i$ was generated from the Bernoulli distribution $B(0.5)$ because we assumed a randomized control trial design in this study.

\paragraph{Observed survival time and censoring indicators}
To obtain the observed survival times and censoring indicators, the true survival and censoring times must be generated first. In the simulation, we referenced the data generation model in Powers et al. (2018)\cite{Powers2018}

and generated the true survival times from the Cox proportional hazards model as follows:
\begin{equation}
h(t|\bm{x}_i, z_i) = h_0(t)\exp(\mu(\bm{x}_i) + (z_i - 0.5)\tau(\bm{x}_i)), \label{true_surv_cox}
\end{equation}
where $\mu(\bm{x}_i)$ is the main effect function and $\tau(\bm{x}_i)$ is the treatment effect function. To evaluate the performance of each method under various conditions, a linear function (Eq.\ref{M1}, Eq.\ref{T1}), a stepwise function (Eq.\ref{M2}, Eq.\ref{T2}), and a nonlinear function (Eq.\ref{M3}, Eq.\ref{T3}) were used for $\tau(\bm{x}_i)$ and $\mu(\bm{x}_i)$. Therefore, the three functions for the main effect were denoted as:
\begin{align}
&\mathrm{M1} : \quad \mu(\bm{x}_i) = 0.5x_{i1} + 0.5x_{i3} + 0.5x_{i5} + 0.5x_{i2} + 0.5x_{i4} - x_{i6}, \label{M1} \\ 
&\mathrm{M2} : \quad \mu(\bm{x}_i) = I(x_{i1} > -1) - I(x_{i3} > 0) + I(x_{i5} > 1) + 0.5x_{i2}x_{i4} - 1.25x_{i6}, \label{M2} \\
&\mathrm{M3} : \quad \mu(\bm{x}_i) = -1.25\sin(x_{i1}x_{i3}) + 2.25/(1 + \exp(-x_{i5})) - 1.5x_{i2}x_{i4}x_{i6} - 1. \label{M3}
\end{align}
The three functions for the treatment effect were denoted as 
\begin{align}
&\mathrm{T1} : \quad \tau(\bm{x}_i) = -x_{i5} - 1.5|x_{i7} + x_{i9}| + 1.5x_{i6} - x_{i8} - x_{i10}, \label{T1} \\ 
&\mathrm{T2} : \quad \tau(\bm{x}_i) = -2I(x_{i5} > -1)I(x_{i7} > 0) - 2I(x_{i7} > 0)I(x_{i9} > 1) - 2.5x_{i6} - x_{i8} + 1.5x_{i10}, \label{T2} \\
&\mathrm{T3} : \quad \tau(\bm{x}_i) = -1.75\sin(x_{i5}x_{i7}) + 3(x_{i5}/(1 + \exp(-x_{i6}x_{i9}))) - 2x_{i8}x_{i9}x_{i10} - 2. \label{T3}
\end{align}
These functions provide nine scenarios for true survival time generation based on various combinations. Additionally, we set the baseline hazard function to $h_0(t) = 2t$. Using the relationship between the survival function and hazard function $-\log(S(t)) = \int_0^t h(s)ds$, Eq.\ref{true_surv_cox} can be transformed into 
\begin{align*}
t = \left\{\frac{-\log(S(t))}{\exp(\mu(\bm{x}_i) + (z_i - 0.5)\tau(\bm{x}_i))}\right\}^{1/2}.
\end{align*}
The set survival rate $S(t) = u \sim U(0,1)$ and true survival time for the $i$-th individual can be calculated as:
\begin{align}
t^*_i = f^*(\bm{x}_i,u_i,z_i) = \left\{\frac{-\log(u_i)}{\exp(\mu(\bm{x}_i) + (z_i - 0.5)\tau(\bm{x}_i))}\right\}^{1/2}. \label{true_surv}
\end{align}
For the censoring time, we assumed the maximum following time $t = 3$ and generated the censoring time using the following functions:
\begin{align*}
f^c(\bm{x}_i) = 1.1\exp(1 - \sin(x_{i1}x_{i3}) + 3(z_i - 0.5)x_{i8} + \varepsilon), \quad \varepsilon \sim N(0,1). 
\end{align*}
The censoring was defined as $c = \min(f^c(\bm{x}_i),3)$. Therefore, under these settings, the observed survival time $t_i$ and censoring indicator $\delta_i$ can be expressed as
\begin{align*}
&t_i = \min(t^*_i, c_i) \\
&\delta_i = I(t^*_i < c_i).
\end{align*}

\paragraph{True HTE}
To evaluate the performance of each method, we defined the true HTE for individual $i$ as the survival rate difference between the control and treatment groups at $t_0 = 2$ (90th-percentile point of the observed survival time)on covariates $\bm{x}_i$, denoted as:
\begin{align*}
\Delta(t_0|\bm{x}_i) & = S(t_0|\bm{x}_i, z_i = 1) - S(t_0|\bm{x}_i,z_i = 0) \\
                     & = P(T_i > t_0|\bm{x}_i, z_i = 1) - P(T_i > t_0|\bm{x}_i, z_i = 0),
\end{align*}
where $T_i$ is a random variable for the true survival time, using values from Eq.\ref{true_surv}. We referred to the true HTE creation process proposed by Cui et al.(2023)\cite{Cui2023}. For each individual $i$, we set its covariates $\bm{x}_i$, sampled the true survival time 100,000 times, and calculated the true HTE as:
\begin{align*}
\Delta(t_0|\bm{x}_i) = \frac{1}{N^*}\sum_{n^*=1}^{N^*}I\left( f^*(\bm{x}_i,u_{n^*},z_i = 1) > t_0\right) - \frac{1}{N^*}\sum_{n^*=1}^{N^*}I\left( f^*(\bm{x}_i,u_{n^*},z_i = 0) > t_0\right),
\end{align*}
where $N^* = 100000$ and $t_0 = 2$.

\paragraph{Performance evaluation}
We compared the performance of our proposed method with several existing survival HTE estimation methods, such as Random survival forest using S-Leaner (rsfs)\cite{kunzel2019}, Enriched random survival forest (rsft)\cite{Lu2018}, Virtual twins (rsfvt)\cite{Foster2011} and Causal survival forest (csf)\cite{Cui2023}. To evaluate the performance of each method, we used four metrics: the \textbf{root mean squared error (RMSE)}, \textbf{absolute relative bias (AbsRbias)}, \textbf{Spearman’s rank correlation}, and \textbf{correct classification rate}. First, we evaluated the prediction accuracy using \textbf{RMSE} and \textbf{AbsRbias}, which are defined as follows:
\begin{align*}
\mathbf{RMSE} &= \sum_{i=1}^N\delta_i\left(\Delta(t_0|\bm{x}_i) - \hat{\Delta}(t_0|\bm{x}_i)\right)^2 \\
\mathbf{AbsRbias} &= \sum_{i=1}^N\ \delta_i \left( \frac{\Delta(t_0|\bm{x}_i) - \hat{\Delta}(t_0|\bm{x}_i)}{\Delta(t_0|\bm{x}_i)}\right),
\end{align*}
where $\Delta(t_0|\bm{x}_i)$ is the true HTE and $\hat{\Delta}(t_0|\bm{x}_i)$ is the estimated HTE. Second, the \textbf{Spearman's rank correlation} between the true HTE $\Delta(t_0|\bm{x}_i)$ and estimated HTE $\hat{\Delta}(t_0|\bm{x}_i)$ was used to evaluate the effectiveness of the method for different individuals. Finally, we used the \textbf{Correct Classification Rate} to evaluate how accurately the estimated HTE reflected the true efficacy of the treatment. It was defined as:
\begin{align*}
\mathbf{Correct \ classification \ rate} = \frac{1}{N}\sum_{i = 1}^N I(sign(\Delta(t_0|\bm{x}_i)) = sign(\hat{\Delta}(t_0|\bm{x}_i))),
\end{align*}
where the $sign()$ function returns 1 if the value in parentheses is positive, -1 if it is negative, and 0 if it is zero. Therefore, a high correct classification rate indicates that the estimated HTE more accurately reflects the true efficacy.

The simulation was performed using the \texttt{R 4.1.2.} programming environment, and all previous methods can be implemented using existing \texttt{R} packages. Random survival forests using S-Leaner, rich random survival forests, and virtual twins were implemented using the \texttt{randomForestSRC} package\cite{Ishwaran2022} with the default hyperparameters, and a causal survival forest was implemented using the \texttt{grf} package\cite{Tibshirani2022} with the default hyperparameters. The hyperparameters of the proposed method were set as follows: the maximum number of trees was $M = 500$, the mean depth of each tree-based learner was $\bar{L} = 2$, the learning rate was $V = 0.01$, and the size of the sample used to train the base learner in each boost step was $\eta = \min(N/2, 100 + 6\sqrt{N})$, where $N$ is the size of the entire training sample.

\subsection{Simulation results}
\begin{figure}[tb]
 \centering
 \includegraphics[width = \linewidth]{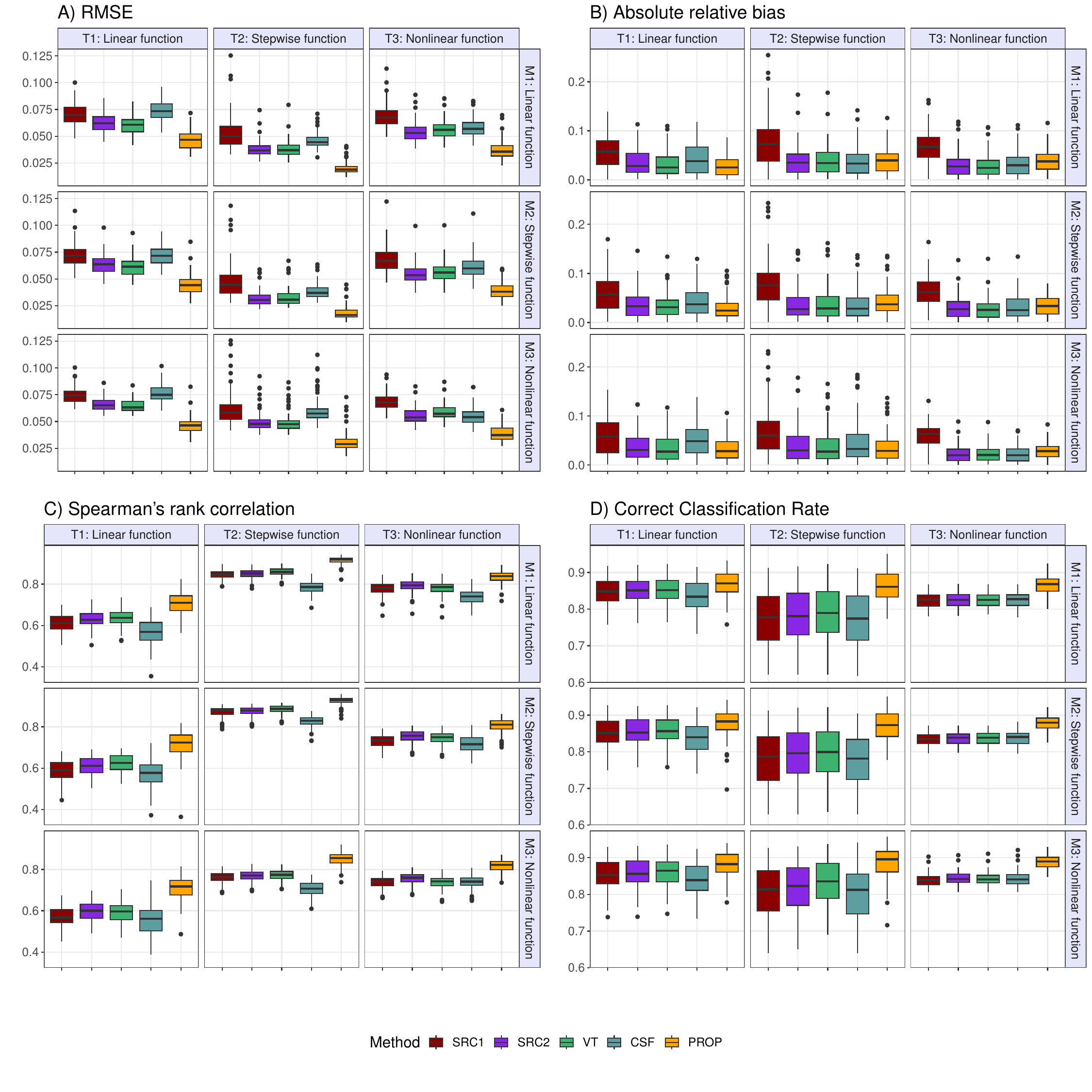}
 \caption{Results of simulations for the nine scenarios. The y-axis is the value of the root mean squared error (RMSE). These boxplots show the RMSE for five different methods, including SRC1 = S-learner strategy using random survival forest, SRC2 = enhancement random survival forest, VT = virtual twin, CSF = causal survival forest, and PROP = proposed method.}
 \label{fig2}
\end{figure}

Here, we present the simulation results for nine scenarios for each method, as illustrated in Fig.\ref{fig2}. The proposed method (PROP) outperformed previous methods in all simulation settings. S-learner using a random survival forest (SRC1) and causal survival forest (CSF) tended to underperform compared to other methods using these simulation datasets. Fig.\ref{fig2} A) provides a detailed presentation of the RMSE for each method. The proposed method consistently exhibited lower RMSE values than the other methods in all scenarios. In particular, the S-learners using a random survival forest (SRC1) generally underperformed. Fig.\ref{fig2} B) displays the relative bias. For clarity, we present the absolute value of the relative bias for each method. When the treatment effects are derived from linear or stepwise functions, the proposed method usually displays a similar or lower relative bias than the other methods. When treatment effects are generated from nonlinear functions, the results of the proposed method are not superior to those of previous methods, but they are comparable. Both the RMSE results and relative bias indicate the superior performance of our proposed method over previous methods, suggesting that it can estimate the HTE more accurately than previous methods. Fig.\ref{fig2} C) shows the Spearman's rank correlation results for each method. For each scenario, the proposed method demonstrated higher correlations than the other methods, indicating that it was capable of capturing the effectiveness of the treatment for each individual. Fig.\ref{fig2} D) presents the correct classification rate based on the estimated HTE for each method. This metric evaluates the diagnostic accuracy of the estimated HTE. Consequently, a high correct classification rate implies that the estimated HTE can diagnose treatment efficacy accurately. Thus, the proposed method outperformed the other methods.

\section{Real data application} 
In this section, we describe how the proposed method applies to a real-world dataset. First, we present a direct interpretation of the results obtained based on the rules and conduct a simple evaluation to assess whether the proposed method correctly interpreted the actual results. Second, we employ a general interpretation approach commonly used in machine learning methods, variable importance, to interpret the results. Third, we demonstrate how the estimated HTE can be used to interpret treatment effectiveness.

We applied our method to a dataset from the AIDS Clinical Trials Group Protocol 175 (ACTG 175)\cite{Hammer1996}, a randomized clinical trial involving HIV1-infected adults with CD4 cell counts between 200 and 500 cells/$mm^3$. The trial included four treatment groups: zidovudine monotherapy, zidovudine plus didanosine combination therapy, zidovudine plus zalcitabine combination therapy, and didanosine monotherapy. The ACTG 175 dataset is available in the R package \texttt{speff2trial}\cite{Juraska2012} and comprises the data of 1762 subjects. In this study, we focused on binary treatment and used a subset of the ACTG 175 dataset that included only two treatment groups: zidovudine monotherapy and zidovudine plus didanosine combination therapy. For simplicity, we refer to zidovudine monotherapy as monotherapy and zidovudine plus didanosine combination therapy as combination therapy. We referred to the 419 subjects who received monotherapy as the control group and the 436 subjects who received combination therapy as the treatment group. The primary endpoint of the ACTG 175 study was the number of days until at least one of the following three events occurred: (i) a decline in the CD4 cell count of at least 50, (ii) progression to AIDS, or (iii) death. We selected 12 covariates for our analysis as proposed by Tsiatis et al. (2008)\cite{Tsiatis2008}. These included five continuous variables: baseline CD4 cell count (cd40; cells/$mm^3$), baseline CD8 cell count (cd80; cells/$mm^3$), age (years), weight (wtkg;kg), and Karnofsky score (karnof; on a scale of 0-100). We also included seven binary variables: hemophilia (hemo; 0 = no, 1 = yes), homosexual activity (homo; 0 = no, 1 = yes), race (0 = white, 1 = other), sex (0 = female, 1 = male), history of intravenous drug use (drugs: 0 = no, 1 = yes), history of antiretroviral therapy (str2; 0 = naive, 1 = experienced), and symptomatic indicators (symptoms: 0 = asymptomatic, 1 = symptomatic). 

The HTE was defined as the difference in three-year (365.25 $\times$ 3 days) survival rates between the treatment and control groups. Therefore, the larger the HTE, the greater the benefit of combination therapy compared to monotherapy. Additionally, as mentioned previously, our proposed method has four hyperparameters: the maximum number of trees, mean depth of each tree-based learner, learning rate, and size of the sample used to train the base learner. We set the hyperparameters to be the same as those used in the simulations.

\subsection{Results of proposed method application}
First, we present the application results for the proposed method and interpret them based on the created rules. The application of the proposed method yielded 16 rules, as shown in Table\ref{table1}.

For each created rule, we provided information about its importance, hazard ratio, and support to help with interpretation. The details of each column are as follows.

\noindent \textbf{Importance}: This column shows the base function importance for each rule (Eq. \ref{lin_imp}). A high base function importance value indicates that the rule is closely related to the HTE. 

\noindent \textbf{Hazard ratio}: This column shows the hazard ratio between the treatment and control groups for each rule. Because our proposed method is based on Cox proportional hazard models, the hazard ratio between the treatment and control groups for each rule is also related to the difference in the survival rate. Thus, we can use the hazard ratio value to interpret the HTE for each rule. A hazard ratio greater than 1 indicates that the HTE is negative and that the control group benefits more than the treatment group. If the hazard ratio is between 0 and 1, it indicates that the HTE is positive, and the treatment group benefits more than the control group. 

\noindent \textbf{Support}: This column shows the support for each rule (Eq. \ref{support}). Support indicates the size of the subgroups, as defined by the rules.
 
To make the results more interpretable and insightful, we selected certain rules from Table\ref{table1} for two reasons. First, the support values for Rules 1, 5, 10, and 12 were less than 0.1. This suggests that the participants in these subgroups constituted less than 10\% of the sample. Consequently, the number of these subgroups may have been too small to provide generalizable results. Second, to obtain more informative results, we focused on the rules which were of greater importance to the HTE. Hence, we finally selected six rules with support values greater than 0.1 and base function importance values exceeding the average value of 26.54. The hazard ratios for these rules fell between 0 and 1, indicating that the subjects within these subgroups benefited more from combination therapy than from monotherapy. The hazard ratios for the first four rules were similar and smaller than those for the last two rules. This suggests that subjects who followed the last two rules benefited less from combination therapy than did those who followed the first four rules. To elucidate this further:

\begin{itemize}
\item The rules ``cd40 $<$ 266.5 \& age $<$ 39.5,’’ ``cd40 $>=$ 298.5 \& age $>=$ 41.5,’’ and ``cd40 $<$ 268 \& age $<$ 39’` suggest that combination therapy is more beneficial than monotherapy for subjects with a baseline CD4 cell count of at least 298.5 who are aged 41.5 years or more, and for those with a baseline CD4 cell count below 266.5 who are aged under 39.5 years. Notably, the subgroups defined by the rules ``cd40 $<$ 266.5 \& age $<$ 39.5’’ and ``cd40 $<$ 268 \& age $<$ 39’’ were nearly overlapping. Consequently, the estimated HTE for subjects who follow these rules would be much greater than those for subjects who follow the other rules. In other words, the subjects who follow these rules can benefit more from the combination therapy than other subjects.

\item The rule ``homo $<$ 0.5 \& race $<$ 0.5 \& cd40 $<$ 359’’ suggests that Caucasians who do not engage in homosexual activity and have baseline CD4 cell counts below 359 benefit more from combination therapy than from monotherapy.

\item The rule ``symptom $<$ 0.5 \& cd80 $>=$ 790.5 \& age $>=$ 28.5’’ points out that asymptomatic subjects, with baseline CD8 cell counts above 790.5 and aged 28.5 years or more, are likely to derive more benefit from combination therapy than from monotherapy.

\item The rule ``wtkg $<$ 97.26 \& cd80 $>=$ 814 \& cd40 $>=$ 335’’ indicates that subjects weighing less than 97.26 kg, with baseline CD8 cell counts of at least 814 and baseline CD4 cell counts of at least 335, also tend to benefit more from combination therapy than from monotherapy.
\end{itemize}

\begin{figure}[tb]
 \centering
 \includegraphics[width = 0.75\linewidth]{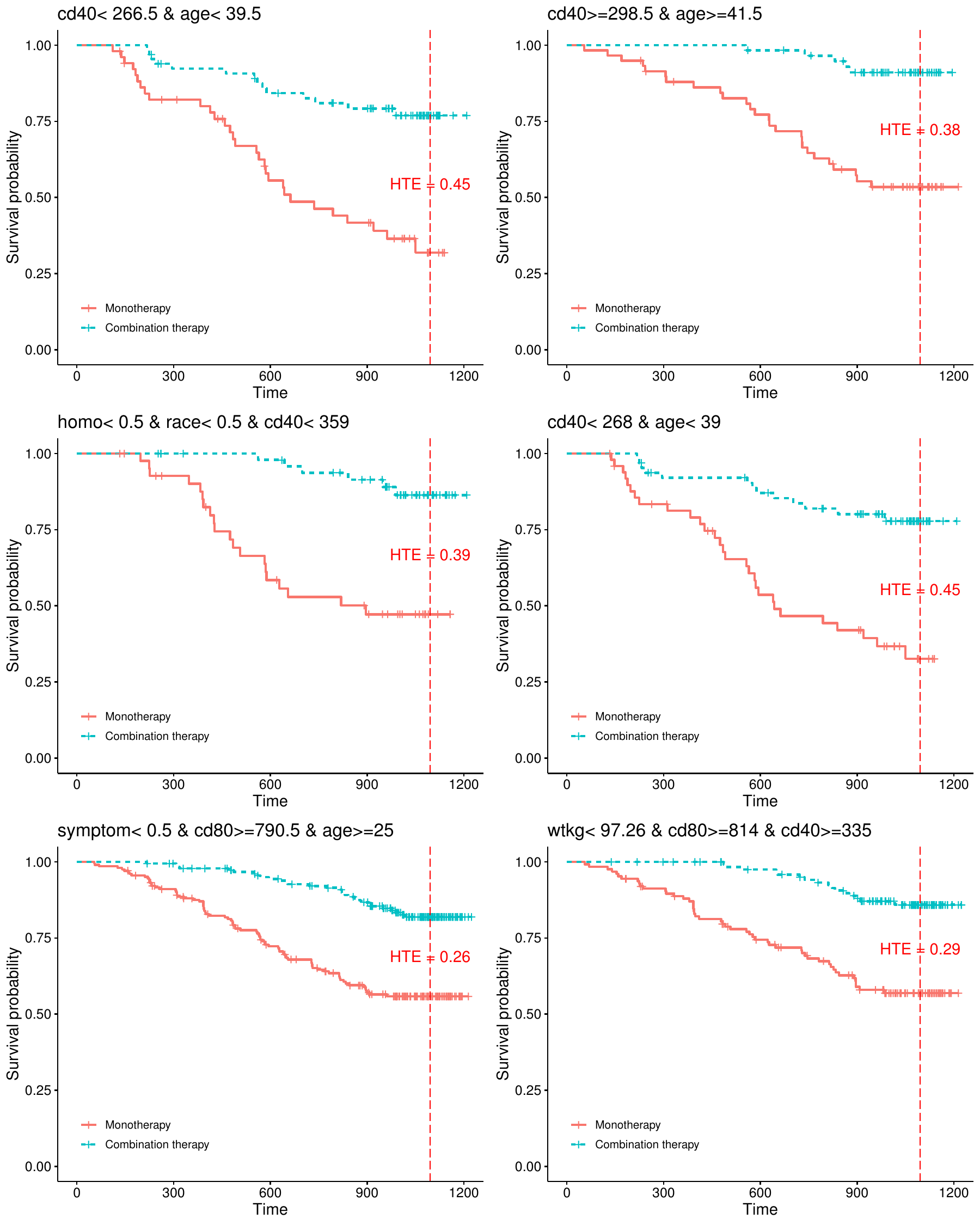}
 \caption{Kaplan--Meier plot for the subgroup defined by each rule. Red curves represent the survival rate for monotherapy, while blue curves represent the survival rate for combination therapy. The red vertical dashed line indicates the 3-year time point. The actual HTE for each subgroup was also calculated}
 \label{km_selected}
\end{figure}

Additionally, we provide a simple evaluation to assess whether the results of the proposed method correctly interpret the actual results. We used the Kaplan--Meier method to calculate the actual HTE for each subgroup, as shown in Fig. \ref{km_selected}. For all the subgroups defined by the selected rules, the effect of combination therapy was superior to that of monotherapy. The HTE for the subgroups ``cd40 $>=$ 298.5 \& age $>=$ 41.5’’ and ``homo $<$ 0.5 \& race $<$ 0.5 \& cd40 $<$ 359’’ were similar, but greater than the HTE for the subgroups ``symptom $<$ 0.5 \& cd80 $>=$ 790.5 \& age $>=$ 28.5’’ and ``wtkg $<$ 97.26 \& cd80 $>=$ 814 \& cd40 $>=$ 335.’’ As for subgroups ``cd40 $<$ 266.5 \& age $<$ 39.5’’ and ``cd40 $<$ 268 \& age $<$ 39,’’ their HTE estimates were higher than those of other subjects. Thus, the interpretations based on the estimated hazard ratio and actual HTE for each rule showed similar trends. This consistency between the two interpretations allows the application of the proposed method to directly identify the subgroups that might derive greater benefits from combination therapy. 

\begin{figure}[tb]
 \centering
 \includegraphics[width = \linewidth]{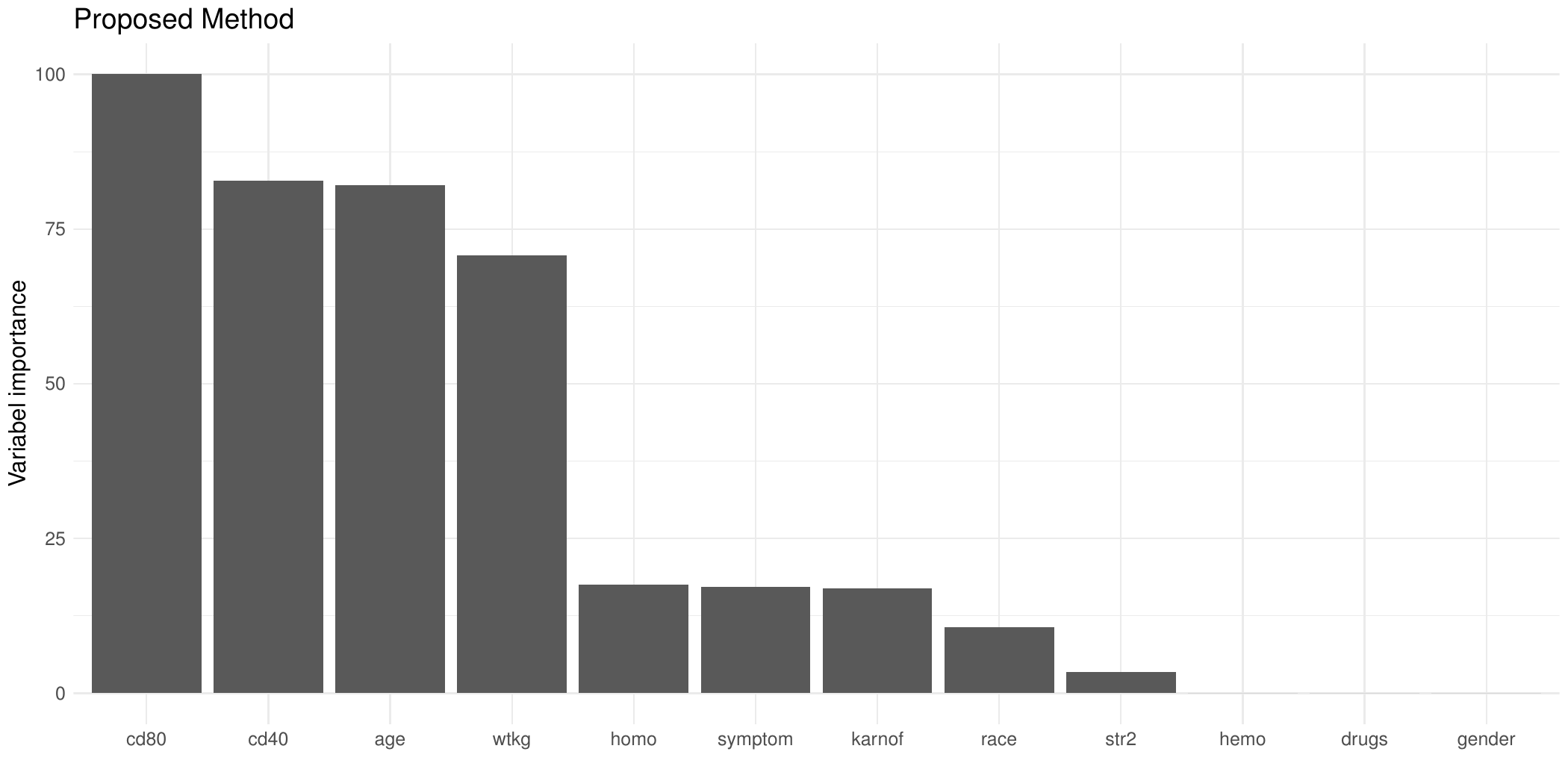}
 \caption{Variable importance of proposed method}
 \label{vip_pic}
\end{figure}

Second, we demonstrated the interpretation of variable importance of the proposed method, as shown in Fig. \ref{vip_pic}. This result indicates that the baseline CD4 cell count, baseline CD8 cell count, weight, and age are much more important than most other variables. The Karnofsky score, homosexual activity, race, history of antiretroviral therapy, and symptomatic indicators contributed less to the HTE. Hemophilia and history of intravenous drug use did not contribute to the HTE. 

Third, we demonstrate how the estimated HTE is utilized to assess the effectiveness of the treatment. We randomly selected 10 subjects from the dataset and estimated their HTE, as shown in Table \ref{table2}. The estimated HTE for Subjects 3, 5, and 7 were 0.44, 0.73, and 0.67, respectively, which were much higher than those for the other subjects. This finding suggests that these subjects would benefit more from combination therapy than from other treatments. By contrast, the estimated HTE values for Subjects 6 and 8 were 0.00 and 0.01, respectively, indicating that these subjects were unlikely to derive any obvious benefit from combination therapy compared to monotherapy.

\begin{figure}[tb]
 \centering
 \includegraphics[width = \linewidth]{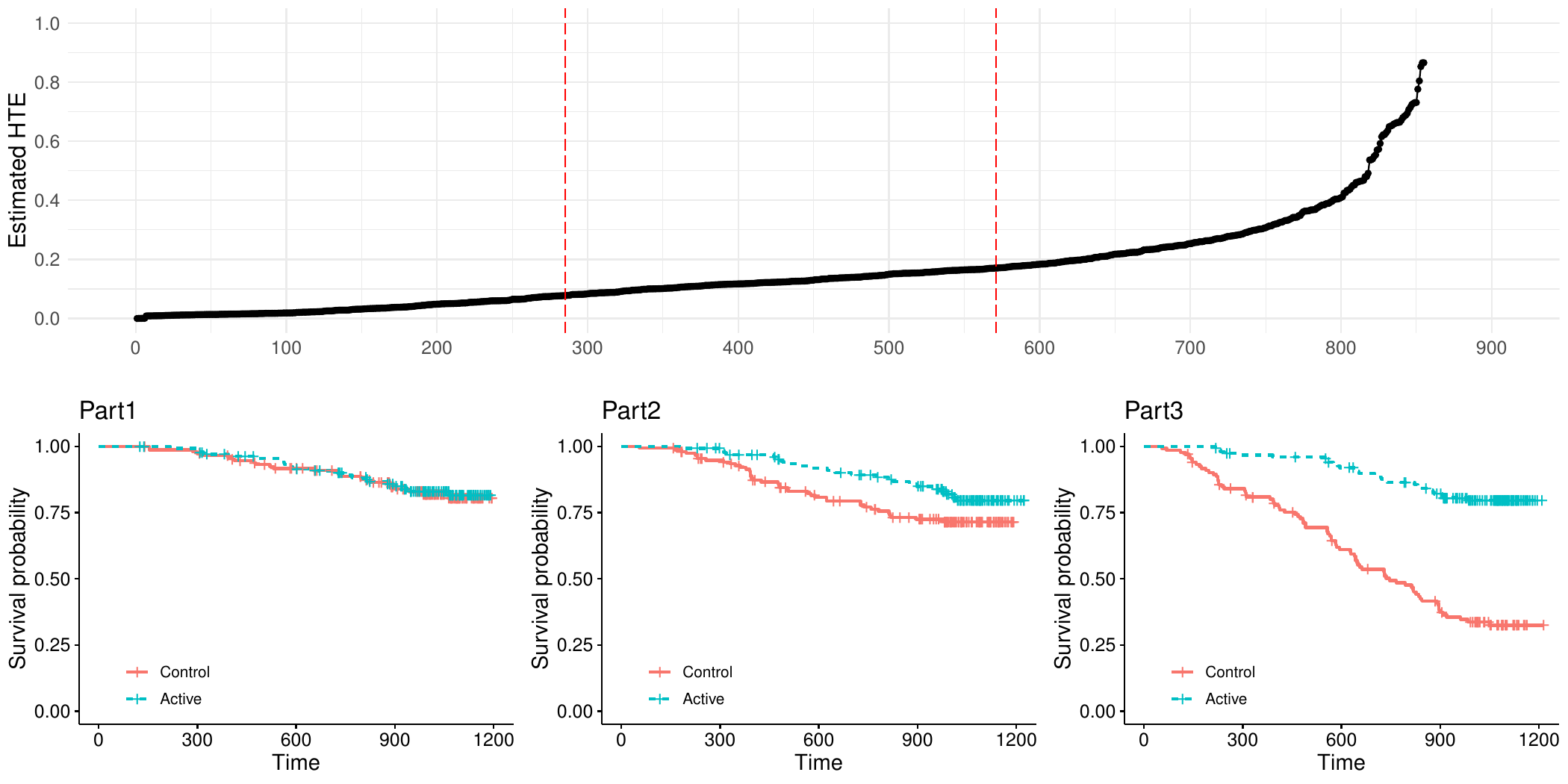}
 \caption{The comparison between the estimated HTE and actual treatment effect. The first row shows the plot of the estimated HTE in ascending order. The horizontal axis represents the order of the estimated HTE for each subject, and the vertical axis displays the corresponding estimated HTE values. The second row displays the Kaplan--Meier plots for the corresponding parts in the first row. The red curve represents the control group and the blue curve represents the treatment group.}
 \label{fig5}
\end{figure}

In addition, we provide an overall interpretation based on the estimated HTE. To achieve this, we ranked the participants based on their estimated HTE and divided them into three equal-sized groups: the low, moderate, and high HTE groups. We subsequently generated a Kaplan--Meier plot for each of these groups, as depicted in Fig\ref{fig5}. These plots suggest that subjects in the low HTE group did not derive much benefit from combination therapy compared to that from monotherapy. By contrast, patients in the high HTE group showed substantial benefits from combination therapy. Therefore, the estimated HTE can be a valuable metric for distinguishing between subjects who are likely to benefit greatly from combination therapy and those who may not experience any pronounced advantages.

\section{Conclusion}
In this study, we introduce an innovative approach to estimate survival HTE based on the survival RuleFit method. The primary strength of the proposed method is its ability to formulate interpretable models, facilitating effortless interpretation of the relationships between HTE and covariates through rule-based terms. Moreover, our approach combines the concepts of S-learner and shared-basis conditional mean regression. This combination serves to avoid bias in HTE estimation caused by differences in the base functions and main effects between the treatment and control group models. Numerous simulations were conducted to assess the predictive performance of the proposed method under diverse conditions. The simulation results indicate that the proposed method outperforms existing approaches, namely, RMSE, Spearman's rank correlation, and correct classification rate. Regarding the absolute relative bias, the proposed method is still comparable to the best-performing method, the CSF. Thus, the simulation results indicate that our proposed method has adequate prediction performance compared to existing methods. The effectiveness of the proposed method is further supported by its application to real data from the HIV study ACTG 175. First, the results of applying the proposed method are presented. Subsequently, some of the most important and general rules are selected to interpret the application results. The interpretation based on the estimated model showed trends similar to the actual results, demonstrating the correctness of the interpretation. Furthermore, we introduced a method to interpret the estimated HTE and compare it to the actual HTE. The results show that the trend of the estimated HTE is similar to that of the actual HTE, indicating that the estimated HTE of the proposed method is correct. 

In summary, the simulation results indicate that the prediction performance of the proposed method is comparable to that of previous methods, whereas real data application shows that the proposed method can correctly interpret actual results. However, in this study, we focused on the RCT dataset, and its application to observational studies also needs to be considered.

\subsection*{Conflict of interest}

No conflicts of interest.

%Bibliography
\bibliographystyle{unsrt}  
\bibliography{references}  

\begin{thebibliography}{10}

\bibitem{Hill2011}
Jennifer~L. Hill.
\newblock Bayesian nonparametric modeling for causal inference.
\newblock {\em Journal of Computational and Graphical Statistics}, 20:217--240,
  1 2011.

\bibitem{Foster2011}
Jared~C. Foster, Jeremy~M.G. Taylor, and Stephen~J. Ruberg.
\newblock Subgroup identification from randomized clinical trial data.
\newblock {\em Statistics in Medicine}, 30:2867--2880, 10 2011.

\bibitem{Tian2014}
Lu~Tian, Lihui Zhao, and L.~J. Wei.
\newblock Predicting the restricted mean event time with the subject's baseline
  covariates in survival analysis.
\newblock {\em Biostatistics}, 15, 4 2014.

\bibitem{Henderson2016}
Nicholas~C. Henderson, Thomas~A. Louis, Chenguang Wang, and Ravi Varadhan.
\newblock Bayesian analysis of heterogeneous treatment effects for
  patient-centered outcomes research.
\newblock {\em Health Services and Outcomes Research Methodology}, 16:213--233,
  12 2016.

\bibitem{Athey2016}
Susan Athey and Guido Imbens.
\newblock Recursive partitioning for heterogeneous causal effects.
\newblock {\em Proceedings of the National Academy of Sciences},
  113:7353--7360, 7 2016.

\bibitem{Wager2018}
Stefan Wager and Susan Athey.
\newblock Estimation and inference of heterogeneous treatment effects using
  random forests.
\newblock {\em Journal of the American Statistical Association},
  113:1228--1242, 7 2018.

\bibitem{Sugasawa2019}
Shonosuke Sugasawa and Hisashi Noma.
\newblock Estimating individual treatment effects by gradient boosting trees.
\newblock {\em Statistics in Medicine}, 38:5146--5159, 11 2019.

\bibitem{Cui2023}
Yifan Cui, Michael~R Kosorok, Erik Sverdrup, Stefan Wager, and Ruoqing Zhu.
\newblock Estimating heterogeneous treatment effects with right-censored data
  via causal survival forests.
\newblock {\em Journal of the Royal Statistical Society Series B: Statistical
  Methodology}, 85:179--211, 5 2023.

\bibitem{Price2018}
W.~Nicholson Price.
\newblock Big data and black-box medical algorithms.
\newblock {\em Science Translational Medicine}, 10, 12 2018.

\bibitem{Petch2022}
Jeremy Petch, Shuang Di, and Walter Nelson.
\newblock Opening the black box: The promise and limitations of explainable
  machine learning in cardiology.
\newblock {\em Canadian Journal of Cardiology}, 38:204--213, 2 2022.

\bibitem{Friedman2008}
Jerome~H. Friedman and Bogdan~E. Popescu.
\newblock Predictive learning via rule ensembles.
\newblock {\em The Annals of Applied Statistics}, 2:916--954, 9 2008.

\bibitem{Fokkema2020}
Marjolein Fokkema.
\newblock Fitting prediction rule ensembles with r package pre.
\newblock {\em Journal of Statistical Software}, 92:1--30, 2020.

\bibitem{kunzel2019}
Sören~R. Künzel, Jasjeet~S. Sekhon, Peter~J. Bickel, and Bin Yu.
\newblock Metalearners for estimating heterogeneous treatment effects using
  machine learning.
\newblock {\em Proceedings of the National Academy of Sciences},
  116:4156--4165, 3 2019.

\bibitem{Powers2018}
Scott Powers, Junyang Qian, Kenneth Jung, Alejandro Schuler, Nigam~H. Shah,
  Trevor Hastie, and Robert Tibshirani.
\newblock Some methods for heterogeneous treatment effect estimation in high
  dimensions.
\newblock {\em Statistics in Medicine}, 37:1767--1787, 5 2018.

\bibitem{Dandl2022}
Susanne Dandl, Andreas Bender, and Torsten Hothorn.
\newblock Heterogeneous treatment effect estimation for observational data
  using model-based forests.
\newblock {\em arXiv:2210.02836}, 10 2022.

\bibitem{Zhu2020}
Jie Zhu and Blanca Gallego.
\newblock Targeted estimation of heterogeneous treatment effect in
  observational survival analysis.
\newblock {\em Journal of Biomedical Informatics}, 107:103474, 7 2020.

\bibitem{ROSENBAUM1983}
Paul~R. Rosenbaum and Donald~B. Rubin.
\newblock The central role of the propensity score in observational studies for
  causal effects.
\newblock {\em Biometrika}, 70:41--55, 1983.

\bibitem{Ridgeway2007}
Greg Ridgeway.
\newblock Generalized boosted models: A guide to the gbm package.
\newblock {\em URL http://cran. open-source-solution.
  org/web/packages/gbm/vignettes/gbm. pdf}, 2007.

\bibitem{Yuan2006}
Ming Yuan and Yi~Lin.
\newblock Model selection and estimation in regression with grouped variables.
\newblock {\em Journal of the Royal Statistical Society: Series B (Statistical
  Methodology)}, 68:49--67, 2 2006.

\bibitem{Lu2018}
Min Lu, Saad Sadiq, Daniel~J. Feaster, and Hemant Ishwaran.
\newblock Estimating individual treatment effect in observational data using
  random forest methods.
\newblock {\em Journal of Computational and Graphical Statistics}, 27:209--219,
  1 2018.

\bibitem{Ishwaran2022}
Hemant Ishwaran and Udaya~B. Kogalur.
\newblock randomforestsrc: Fast unified random forests for survival,
  regression, and classification (rf-src).
\newblock {\em R package version 3.1.1}, 2022.

\bibitem{Tibshirani2022}
Julie Tibshirani, Susan Athey, Erik Sverdrup, and Stefan Wager.
\newblock grf: Generalized random forests.
\newblock {\em R package version 2.2.0}, 2022.

\bibitem{Hammer1996}
Scott~M. Hammer, David~A. Katzenstein, Michael~D. Hughes, Holly Gundacker,
  Robert~T. Schooley, Richard~H. Haubrich, W.~Keith Henry, Michael~M. Lederman,
  John~P. Phair, Manette Niu, Martin~S. Hirsch, and Thomas~C. Merigan.
\newblock A trial comparing nucleoside monotherapy with combination therapy in
  hiv-infected adults with cd4 cell counts from 200 to 500 per cubic
  millimeter.
\newblock {\em New England Journal of Medicine}, 335:1081--1090, 10 1996.

\bibitem{Juraska2012}
Michal Juraska, Peter~B. Gilbert, Xiaomin Lu, Min Zhang, Marie Davidian, and
  Anastasios~A. Tsiatis.
\newblock speff2trial: Semiparametric efficient estimation for a two-sample
  treatment effect.
\newblock {\em R package version 1.0.4}, 2012.

\bibitem{Tsiatis2008}
Anastasios~A. Tsiatis, Marie Davidian, Min Zhang, and Xiaomin Lu.
\newblock Covariate adjustment for two-sample treatment comparisons in
  randomized clinical trials: A principled yet flexible approach.
\newblock {\em Statistics in Medicine}, 27:4658--4677, 10 2008.

\end{thebibliography}

\clearpage

\clearpage

%\begin{threeparttable}[t]
\begin{table}[tb]
\centering
\caption{Results of the proposed method}
\label{table1}
\begin{tabular}{ll|cccc}
\hline
Rules&       & Importance & Harzard\_Ratio & Support \\ \hline
Rule 1 & wtkg\textgreater{}=80.63 \& cd80\textgreater{}=1524                           & 100.00     & 0.21           & 0.04  \\
\textbf{Rule 2\tnote{*}}& \textbf{cd40\textless 266.5 \& age\textless 39.5}                                      & \textbf{39.05}      & \textbf{0.71}   & \textbf{0.14}  \\
\textbf{Rule 3\tnote{*}}& \textbf{cd40\textgreater{}=298.5 \& age\textgreater{}=41.5}                            & \textbf{36.93}      & \textbf{0.72}  & \textbf{0.14}  \\
\textbf{Rule 4\tnote{*}}& \textbf{homo\textless 0.5 \& race\textless 0.5 \& cd40\textless 359}                   & \textbf{34.63}      & \textbf{0.71}  & \textbf{0.11}  \\
Rule 5& cd80\textgreater{}=817 \& age\textgreater{}=48.5                              & 32.54      & 0.61    & 0.04    \\
\textbf{Rule 6\tnote{*}}& \textbf{cd40\textless 268 \& age\textless 39}                                          & \textbf{32.42}      & \textbf{0.75}           & \textbf{0.13}   \\
\textbf{Rule 7\tnote{*}}& \textbf{symptom\textless 0.5 \& cd80\textgreater{}=790.5 \& age\textgreater{}=25}      & \textbf{30.87}      & \textbf{0.83}           & \textbf{0.46}   \\
\textbf{Rule 8\tnote{*}}& \textbf{wtkg\textless 97.26 \& cd80\textgreater{}=814 \& cd40\textgreater{}=335}       & \textbf{30.12}      & \textbf{0.82}           & \textbf{0.30}  \\
Rule 9& symptom\textless 0.5 \& cd80\textgreater{}=790.5 \& age\textgreater{}=28.5    & 24.82      & 0.86           & 0.39    \\
Rule 10& homo\textgreater{}=0.5 \& cd40\textgreater{}=380.5 \& cd80\textgreater{}=1353 & 22.44      & 0.74           & 0.05    \\
Rule 11& wtkg\textgreater{}=75.05 \& karnof\textgreater{}=95                           & 18.88      & 0.88           & 0.29    \\
Rule 12& str2\textless 0.5 \& wtkg\textgreater{}=78.25 \& cd40\textless 322.5          & 11.27      & 0.87           & 0.07    \\
Rule 13& cd40\textless 268 \& karnof\textgreater{}=85 \& cd80\textgreater{}=647        & 7.87       & 0.93           & 0.12    \\
Rule 14& wtkg\textless 97.26 \& cd80\textgreater{}=814                                 & 6.85       & 0.96           & 0.55    \\
Rule 15& karnof\textgreater{}=85                                                       & 6.33       & 0.91           & 0.96    \\
Rule 16& cd40\textless 415.5                                                           & 0.17       & 1.00           & 0.78    \\ \hline
\end{tabular}
\end{table}
%%% footnot %%%
%\begin{tablenotes}%[para,flushleft,online,normal] %(default:normal)
%\item[\textbf{*}] Rules selected for interpreting the results of the proposed method %application.
%\end{tablenotes}
%\end{threeparttable}

\begin{table}[tb]
\caption{Estimated HTE for 10 randomly chosen subjects}
\label{table2}
\begin{tabular}{c|c|cccccccccccc}
\hline
No. & HTE  & cd40 & cd80 & age & wtkg  & karnof & hemo & homo & drugs & race & gender & str2 & symptom \\ \hline
1   & 0.06 & 458  & 1861 & 16  & 41.00 & 100    & 1    & 0    & 0     & 0    & 1      & 1    & 0       \\
2   & 0.13 & 319  & 1412 & 37  & 79.38 & 100    & 1    & 0    & 0     & 0    & 1      & 0    & 1       \\
3   & 0.44 & 257  & 960  & 23  & 87.00 & 100    & 0    & 0    & 0     & 0    & 0      & 0    & 0       \\
4   & 0.09 & 390  & 780  & 52  & 90.58 & 90     & 1    & 0    & 0     & 0    & 1      & 1    & 0       \\
5   & 0.73 & 418  & 1722 & 30  & 81.30 & 90     & 0    & 1    & 0     & 0    & 1      & 0    & 0       \\
6   & 0.00 & 255  & 578  & 46  & 67.70 & 80     & 0    & 1    & 0     & 0    & 1      & 1    & 1       \\
7   & 0.67 & 428  & 1832 & 28  & 90.50 & 100    & 0    & 1    & 0     & 1    & 1      & 1    & 0       \\
8   & 0.01 & 337  & 699  & 34  & 62.60 & 100    & 0    & 1    & 0     & 0    & 1      & 0    & 0       \\
9   & 0.16 & 310  & 1068 & 44  & 82.70 & 90     & 0    & 1    & 0     & 0    & 1      & 0    & 1       \\
10  & 0.08 & 297  & 501  & 36  & 62.20 & 90     & 0    & 0    & 0     & 0    & 0      & 1    & 0       \\ \hline
\end{tabular}
\end{table}

\end{document}